\documentclass{article}
\usepackage{epsfig}

\date{\today}

\usepackage{amsmath}
\usepackage{amssymb}
\usepackage[hang,nooneline,scriptsize]{subfigure}
\renewcommand{\thefootnote}{\arabic{footnote}}

\makeatletter
\let\@fnsymbol\@arabic
\makeatother

\begin{document}

\title{Rotating Boson Stars in Einstein-Gauss-Bonnet gravity}

\author{{\large Yves Brihaye 
}$^{\dagger}$\footnote{email: yves.brihaye@umons.ac.be} , 
{\large J\"urgen Riedel }$^{\ddagger}$\footnote{email:
jriedel@thescienceinstitute.com}
\\ \\
$^{\dagger}${\small Physique-Math\'ematique, Universit\'e de
Mons, 7000 Mons, Belgium}\\ 
$^{\ddagger}${ \small Institut f\"ur Physik, Universit\"at Oldenburg, 26111 Oldenburg, Germany}  }

\date{\today}
\setlength{\footnotesep}{0.5\footnotesep}
\newcommand{\dd}{\mbox{d}}
\newcommand{\tr}{\mbox{tr}}
\newcommand{\la}{\lambda}
\newcommand{\ka}{\kappa}
\newcommand{\f}{\phi}
\newcommand{\vf}{\varphi}
\newcommand{\F}{\Phi}
\newcommand{\al}{\alpha}
\newcommand{\ga}{\gamma}
\newcommand{\de}{\delta}
\newcommand{\si}{\sigma}
\newcommand{\bomega}{\mbox{\boldmath $\omega$}}
\newcommand{\bsi}{\mbox{\boldmath $\sigma$}}
\newcommand{\bchi}{\mbox{\boldmath $\chi$}}
\newcommand{\bal}{\mbox{\boldmath $\alpha$}}
\newcommand{\bpsi}{\mbox{\boldmath $\psi$}}
\newcommand{\brho}{\mbox{\boldmath $\varrho$}}
\newcommand{\beps}{\mbox{\boldmath $\varepsilon$}}
\newcommand{\bxi}{\mbox{\boldmath $\xi$}}
\newcommand{\bbeta}{\mbox{\boldmath $\beta$}}
\newcommand{\ee}{\end{equation}}
\newcommand{\eea}{\end{eqnarray}}
\newcommand{\be}{\begin{equation}}
\newcommand{\bea}{\begin{eqnarray}}
\newcommand{\varA}[1]{{\operatorname{#1}}}

\newcommand{\ii}{\mbox{i}}
\newcommand{\e}{\mbox{e}}
\newcommand{\pa}{\partial}
\newcommand{\Om}{\Omega}
\newcommand{\vep}{\varepsilon}
\newcommand{\bfph}{{\bf \phi}}
\newcommand{\lm}{\lambda}
\def\theequation{\arabic{equation}}
\renewcommand{\thefootnote}{\fnsymbol{footnote}}
\newcommand{\re}[1]{(\ref{#1})}
\newcommand{\R}{{\rm I \hspace{-0.52ex} R}}
\newcommand{\N}{{\sf N\hspace*{-1.0ex}\rule{0.15ex}%
{1.3ex}\hspace*{1.0ex}}}
\newcommand{\Q}{{\sf Q\hspace*{-1.1ex}\rule{0.15ex}%
{1.5ex}\hspace*{1.1ex}}}
\newcommand{\C}{{\sf C\hspace*{-0.9ex}\rule{0.15ex}%
{1.3ex}\hspace*{0.9ex}}}
\newcommand{\eins}{1\hspace{-0.56ex}{\rm I}}
\renewcommand{\thefootnote}{\arabic{footnote}}
 \maketitle
\begin{abstract} 
A self-interacting SU(2)-doublet of complex scalar fields,  minimally coupled to Einstein-Gauss-Bonnet
gravity is considered in five space-time dimensions. The classical equations admit two families of solitons
corresponding to spinning and non-spinning bosons stars. The generic solutions are constructed numerically
and agree with exact results that are available in special limits of the parameters.
 The pattern of the boson stars is 
shown to be qualitatively affected by the Gauss-Bonnet coupling constant.
\end{abstract}

\section{Introduction}
Boson stars and Q-balls have been known for a long time. They are
non-topological solitons \cite{fls,lp} characterized by a conserved Noether
charge associated with a U(1) symmetry of the Lagrangian. 
One of the first constructions of  non-topological solitons
was achieved in \cite{coleman}, within a field theory describing a self-interacting complex scalar field.
Q-balls are classical solutions; they are stationary with an explicite
time-dependent phase with frequency $\omega$. The conserved Noether charge  $Q$
is then related to the global phase invariance of the theory and is directly proportional
to the frequency; $Q$ can further be interpreted as the particle number. 
In \cite{vw}, it was shown
that a non-normalizable $\Phi^6$-potential is necessary to support soliton solutions.
Using such a potential, several $Q$-ball solutions in $3+1$ dimensions have been studied in details in 
\cite{vw,kk1,kk2}. 

The interest for these type of classical solutions was enhanced 
in particular after it was shown \cite{kusenko,dm} that supersymmetric extensions of the
Standard Model (SM)  also possess $Q$-ball solutions.
 In \cite{ct}  an effective  potential involving these effects
 was suggested and the
properties of the corresponding $Q$-balls have been investigated \cite{cr}.  
Many astrophysical implications have been discussed, see \cite{implications} for a non exhaustive list.

Recently several authors addressed the existence and the construction of globally regular
solutions in scalar field theories coupled (minimally or not) to gravity in more than four dimensions
\cite{Astefanesei:2003qy},\cite{Prikas:2004fx},\cite{Hartmann:2012gw}. 
Several features of the four-dimensional boson stars hold also in higher dimensions. For example the
spectrum of the solutions as functions of the frequency $\omega$ present several branches existing on
specific intervals of this parameter. Also, sequences of radially excited solutions exist.


Perhaps, one of the most promising way to discover new features of these solitons is to emphasize
their spinning properties.
Classical solutions living in d-dimensions can have $(d-1)/2$
independent angular momenta. It is therefore natural to emphasize the construction of
rotating boson stars and Q-balls.
The first construction of rotating boson stars in five dimensional is reported in  \cite{Hartmann:2010pm}. In this paper it is shown that rotating solitons in d=5
 can be accommodated if two complex scalar fields are present, assuming the two angular momenta to
 be equal, the classical equations leads to a system of ordinary differential equations.
 
 
 With the motivation of constructing black holes with only one Killing field, the
 type of field configurations  proposed in \cite{Hartmann:2010pm} was reconsidered
 in \cite{Dias:2011at}.  A cosmological constant was supplemented to the model 
 and solitons as well as hairy black hole solutions have been constructed numerically.
 Due to the choice of a negative cosmological constant, these  asymptotically ADS
 solutions exist without any potential for the scalar field.
 The extension of such objects to higher dimensions was emphasized in \cite{Stotyn:2013yka}.

For $d>4$, the standard Einstein-Hilbert action is not 
the only possibility for modelling gravity. 
It is well known that a hierarchy of extended lagrangians exist while increasing the number of dimensions, say $d$.
In particular for $d=5$, a Gauss-Bonnet term can be added to the Einstein-Hilbert action.
It is therefore natural to study the effects of  the Gauss-Bonnet term to the spectrum of boson stars.
The non-spinning boson stars were constructed  with Einstein-Gauss-Bonnet action in \cite{Hartmann:2013tca}. It 
is the purpose of this paper to extend this results for spinning solutions.

The paper is organized as follows. In section~2, we present the model, the general field equations
and the ansatz allowing us to reduce these equations to a system of differential equations. 
Some exact results, including the asymptotic forms of the generic solutions, 
are presented in Sect. 3.
New features of the non-spinning solutions are pointed out in section~4. 
The results for spinning solutions are presented in sections~5 and  6 respectively in the case of
a null and negative cosmological constant. Section 7 contains conclusions.

\section{The model}

In this paper, we  study Q-balls and boson stars for a self-interacting doublet of scalar fields 
 in $(4+1)$-dimensional space-time and coupled to Einstein-Gauss-Bonnet gravity. 
The action reads~:
\begin{equation}
S= \frac{1}{16\pi G} \int d^5 x \sqrt{-g} \left(R -2\Lambda + 
\frac{\alpha}{2}\left(R^{MNKL} R_{MNKL} - 4 R^{MN} R_{MN} + R^2\right) + 16\pi G {\cal L}_{\rm matter}\right) \ ,
\end{equation}
where $\Lambda=-6/\ell^2$ is the cosmological constant, $\alpha$ the Gauss--Bonnet coupling
and $M,N,K,L \in \{0,1,2,3,4\}$. 
The Lagrangian density for the matter fields ${\cal L}_{\rm matter}$ reads~:
\begin{equation}
{\cal L}_{\rm matter}=  - 
\left(\partial_M\Phi\right)^{\dagger} \partial^M \Phi - U( |\Phi|)  \ \ ,
\end{equation}
where the scalar field $\Phi$ denotes a doublet of scalar fields~: $\Phi = (\phi_1,\phi_2)^t$.
Along with many authors, \cite{ct,cr}, we choose the potential 
\be
\label{susy_pot}
        U( |\Phi|) = m^2 \eta_{susy}^2 \bigg(1 - \exp( - \frac{|\Phi|^2}{\eta_{susy}^2} ) \bigg) 
\ee
which is constructed in such a way as
 to encode some relevant features of supersymmetric extensions of the standard model \cite{ct}. The parameter $m$
denotes the mass of the scalar boson while $\eta_{susy}$ is related to the energy scale below which
supersymmetry is broken.

The coupled field equations for matter and gravity  are obtained from the variation of the
action with respect to the matter and metric fields respectively, leading to
\begin{equation}
\label{full_eqs}
\partial^M  \partial_M \Phi = \frac{\partial U}{\partial \Phi} \ , \  
G_{MN} + \Lambda g_{MN} + \frac{\alpha}{2} H_{MN}=8\pi G T_{MN} \ ,
\end{equation}
where $H_{MN}$ is given by
\begin{equation}
 H_{MN}= 2\left(R_{MABC}R_N^{ABC} - 2 R_{MANB}R^{AB} - 2 R_{MA}R^{A}_N + R R_{MN}\right)
- \frac{1}{2} g_{MN} \left(R^2 - 4 R_{AB}R^{AB} + R_{ABCD} R^{ABCD}\right)
\end{equation}
and $T_{MN}$ is the energy-momentum tensor
\begin{equation}
 T_{MN}=g_{MN} {\cal L}_{\rm matter} - 2\frac{\partial {\cal L}_{\rm matter}}{\partial g^{MN}} \ .
\end{equation}

Our aim is to construct gravitating, rotating  solitons of the above equations. 
In general, such a solution
would possess two independent angular momenta associated to the two orthogonal planes of 
rotation.
Here we  restrict to the case of equal angular momenta, 
the relevant   ansatz for the metric reads \cite{kunz1}
\begin{eqnarray}
\label{metric}
ds^2 & = & -b(r) dt^2 + \frac{1}{f(r)} dr^2 + g(r) d\theta^2 + h(r)\sin^2\theta \left(d\varphi_1 - 
W(r) dt\right)^2 + h(r) \cos^2\theta\left(d\varphi_2 -W(r)dt\right)^2 \nonumber \\
&+& 
\left(g(r)-h(r)\right) \sin^2\theta \cos^2\theta (d\varphi_1 - d\varphi_2)^2 \ ,
\end{eqnarray}
where $\theta$ runs from $0$ to $\pi/2$, while $\varphi_1$ and $\varphi_2$ are 
in the range $[0,2\pi]$.
The corresponding space-times  possess two rotation planes at $\theta=0$ and $\theta=\pi/2$ and the isometry
group is $\mathbb{R}\times U(2)$.
The metric above still leaves the diffeomorphisms related to the definitions of the radial variable $r$ unfixed; 
for the numerical construction, we will fix this  freedom by choosing $g(r)=r^2$. 
The metric (\ref{metric}) then leads to a consistent system
of differential equations \cite{Hartmann:2010pm} when the doublet of 
scalar fields is of the form
\be
          \Phi = \phi(r) e^{i \omega t} \hat \Phi
\ee
where $\hat \Phi$ is a doublet of unit length  depending on the angles only. It needs to be fixed 
appropriately for static or rotating solutions. For static solutions, we have $\hat \Phi = (1,0)^t$ while for
rotating solutions the form 
\be
\hat \Phi = (\sin \theta e^{i \varphi_1},\cos \theta e^{i \varphi_2}  )^t
\ee  
  guarantees the consistency of the ansatz.
The general field-equations are  transformed into a set
of  five differential equations for the functions $f(r), b(r), h(r), W(r)$ and $\phi(r)$

These equations depend on five independent constants: the Newton's constant $G$, 
the cosmological constant $\Lambda$ (or the Anti-de Sitter radius $L$), the Gauss-Bonnet parameter $\alpha$,
the mass of the scalar field $m$, and  the parameter $\eta_{susy}$.

In the case $\Lambda=0$,
the parameters $\eta_{susy}$  can be absorbed in a redefinition of the scale of the scalar
field  while the mass parameter can be absorbed in a rescaling of the space-time coordinates $x_M$.
With these conventions and  the rescaling, the fields equations depend on two dimensionless parameters~:
the Gauss-Bonnet parameter $\alpha$ and $\kappa \equiv 16 \pi G \eta_{susy}^2$.  
In the case $\Lambda < 0$, the boson stars exist even in the absence of a potential ($m=0$).
An appropriate rescaling can then be used to set $\ell$ and $\kappa$ to particular values.

It might be useful to contrast the effective (or 'reduced') matter-lagrangian densities 
$\sqrt{-g} {\cal L}_{matter}$ 
 for the non-spinning and spinning    cases.
In the non-spinning case, the field-equations further imply $h=r^2$, $W(r) = 0$; the effective 
lagrangian for the matter field simplifies to
\be
   {\cal L}_{eff} = r^3 \sin \theta \cos \theta \sqrt{\frac{b}{f}} \left( - U(\phi) - f(\phi')^2 + \omega^2 \frac{\phi^2}{b} \right) 
\ee
In the spinning case, the corresponding formula is slightly more involved~:
\be
   {\cal L}_{eff} = r^2 \sin \theta \cos \theta \sqrt{\frac{b h}{f}} 
   \left( - U(\phi) - f(\phi')^2 + \omega^2 \frac{\phi^2}{b} - \phi^2(\frac{1}{h} + \frac{2}{g})
   - \frac{W^2}{b} - 2 \frac{\omega W}{b} \right) 
\ee
The  effective action densities for the Einstein-Hilbert and Gauss-Bonnet terms
for the metric ansatz (\ref{metric}) can be found e.g. in \cite{Brihaye:2008kh}.

\section{Exact solutions and asymptotics}
As pointed out already, the ansatz above transforms the field equations (\ref{full_eqs})
into a set of five ordinary differential equations. 
In order to provide physically meaningful solutions,  suitable conditions have to be
imposed at $r=0$ (regularity at the origin) and asymptotically (localized solution in flat or AdS space-time). 
Only for specific limits of the coupling constants do the solutions admit a closed form.
In this section we present a few exact results  and the behaviour of the solution for 
$r\rightarrow 0$ and $r\to \infty$ in the generic case. 

\subsection{Asymptotics}
As we want to construct solutions that are 
regular and localized in space and  asymptotically flat, 
several appropriate boundary conditions should be imposed. 
For $r \to \infty$, the metric should  approach Minkowski space-time and the matter fields should vanish.
This   implies the  following behaviour of the five radial functions~:
 \be
 f(r) = 1 + \frac{{\cal U}}{r^2} + o(\frac{1}{r^4}) \ \ , \ \ b(r) = 1 + \frac{{\cal U}}{r^2} + o(\frac{1}{r^4}) 
 \ee
 \be
 h(r) = r^2(1 + o(\frac{1}{r^4})) \ \ , \ \ w(r) = \frac{{\cal W}}{r^4} \ \ , \ \ 
 \phi(r) \sim \frac{1}{r^{3/2}} \exp{- r \sqrt{m^2-\omega^2}} 
 \ee
The parameters ${\cal U,W}$ in the asymptotic expansion are 
used to determine the mass $M$ and angular momentum $J$ of the solutions~: 
\begin{equation}
 M =  - 3 \frac{V_3}{16 \pi G}  {\cal U} \ \ , \ \ 
J  =\frac{V_3}{ 8 \pi G} {\cal W} \ \ , \ \ V_3 = 2 \pi^2 \ .
\end{equation}

\subsection{Taylor expansion}
In order to study the behaviour of the solutions around the origin,  a Taylor
expansion of the solutions can be done. We report it only in the non rotating case, 
(the expressions become  lengthy and not illuminating in the non-rotating case). 
For convenience we use $A^2 \equiv b/f$ and let 
\be
      f(r) = 1 + f_2 r^2 + f_4 r^4 \ \ , \ \ A = A_0 + A_2 r^2 + A_4 r^4 \ \ , \ \ \phi(r) = F_0 + F_2 r^2 + F_4 r^4 + \dots
\ee
Then expanding the field equations in powers of $r$ imply that all coefficients are determined
in terms of  $F_0$ and $A_0$. In particular, for the quadratic terms, we find
\be
      f_2 = \frac{1}{6 \alpha A_0} \left(3 A_0 \pm \sqrt{3} \sqrt{3 A_0^2 + 2 \kappa \alpha(A_0^2 U(f_0)+ \omega^2 F_0^2)}\right)
\ee
\be
       A_2 = \frac{\kappa \omega^2 F_0^2 }{3 A_0(1-2\alpha f_2)} \ \ , \ \ F_2 = \frac{U'(F_0)- 2 F_0 \omega^2}{16 A_0^2}
\ee
The expansion suggests that two solutions should exist for generic values of $\alpha$; they are distinguished
  by the arbitrary sign $\pm$ occuring in   $f_2$. Only the solution corresponding to the 'minus sign' however
possesses a  regular limit for $\alpha \to 0$.  Only this branch  of solutions will be emphasized in the next section.

\subsection{Linearized equation: $\Lambda=0$}
We note that for $\kappa = 0$, the Einstein-Gauss-Bonnet equations only allow regular solutions in Minkowski space-time  (irrespectively of the  Gauss-Bonnet coupling constant).
In the  limit  $\kappa=0$ and assuming for the potential  of a mass term only 
(the mass we note $m$ in this section),  global 
analytical  solutions  can be found for the underlying Klein-Gordon equation. 
 In the non spinning case, the relevant equation for the scalar field reads
 \be
      \phi'' + \frac{3}{r} \phi' + \left(\omega^2 - m^2\right) \phi = 0
 \ee
 which can be solved in terms of the modified Bessel functions $I_1$,$K_1$
  for $0 < \omega < m$~:
 \be 
 \label{bessel_1}
     \phi(r) = \phi_0 \frac{1}{r} I_1\big( \sqrt{m^2-\omega^2} r\big) \ \ , \ \ {\rm for} \ \  r \ll 1 \ \ ; \ \ 
     \phi(r) = \tilde \phi_0 \frac{1}{r} K_1\big( \sqrt{m^2-\omega^2} r\big) \ \ , \ \ {\rm for} \ \ r \gg  1
 \ee
The function $I_1(r)/r$ is finite at the origin and $K_1$ present the desired asymptotic behaviour.
In the spinning case, the equation reads \\
 \be
      \phi'' + \frac{3}{r} \phi' + \left(-\frac{3}{r^2} + \omega^2 - m^2\right) \phi = 0
 \ee
 which can be solved in terms of the modified Bessel functions $I_2$, $K_2$~: 
 \be 
 \label{bessel_2}
     \phi(r) = \phi_0 \frac{1}{r} I_2\big( \sqrt{m^2-\omega^2} r\big) \ \ , \ \ {\rm for} \ \ r \ll 1 \ \ ; \ \ 
     \phi(r) = \tilde \phi_0 \frac{1}{r} K_2\big( \sqrt{m^2-\omega^2} r\big) \ \ , \ \ {\rm for} \ \ r \gg  1
 \ee
 which also   fulfil the relevant boundary conditions.
Solving the equation with the full interacting potential  leads to a global solution
approaching the above functions for $r \ll 1$ and $r \gg 1$
 and fixes the constant $\phi_0, \tilde \phi_0$ in function of $\omega$.
Similar results are discussed in \cite{Copeland:2009as} 
for generic values  of the dimension of space-time,
for different potentials in one scalar field. 

\subsection{Probe limit: $U=0$, $\Lambda<0$}
In the presence of a (negative) cosmological constant Q-balls and boson stars exist in the
absence of a potential; we assume $U=0$ in this section.
In the probe limit (i.e. with $\kappa=0$), the relevant solution to the Einstein-Gauss-Bonnet equations
is the AdS space-time with a modified AdS radius~:
\be~
\label{ads}
            f(r) = b(r) = 1 + \frac{r^2}{\ell_c^2} \ \ , \ \ h(r) = r^2 \ \ , \ \     
            \frac{1}{\ell_c^2} \equiv \frac{1}{\alpha} \left[1 - \sqrt{1 - \frac{2 \alpha}{\ell^2}}\right] \ .
\ee
In particular, the Gauss-Bonnet parameter is bounded: $\alpha \leq \ell^2/2$.
In  this background, the Klein-Gordon equation  leads to the differential equation 
\be
      (r^3 f \phi')' - r^3  \left(\frac{p}{r^2} - \frac{\omega^2}{f}\right) \phi = 0 \ \ , 
      \ \ p=0 : \varA{non-spinning} \ \ , \ \ p=3 : {\rm spinning} 
\ee
which can be solved in  terms of Hypergeometric functions.
For non-spinning soliton, we find
\be
        \phi(r) = \frac{c_0\ell_c^4}{(r^2 + \ell_c^2)^2}
        \phantom{A}_2F_1(\frac{4- \omega \ell_c}{2}, \frac{4+ \omega \ell_c}{2} ; 3 , \frac{\ell_c^2}{r^2 + \ell_c^2} )
\ee
The corresponding Hypergeometric function $\phantom{A}_2F_1(a,b;c,z)$ is divergent in the limit $z \to 1$
(see \cite{magnus}  p. 37) for generic values of $a,b$. However it reduces to a polynom if $a$ or $b$
is a negative integer. As a consequence, the physical relevant solutions correspond to $\omega\ell_c = 4 + 2 k$
for any positive integer $k$. For $k=0$, the Hypergeometric function is a constant and the solution
clearly obeys the boundary conditions of the "fundamental" boson star. The higher values of $k$ lead to
radial excited solutions where $\phi(r)$ present nodes at intermediate values of $r$.

In the case of spinning soliton, we have \cite{Dias:2011at} 
\be
        \phi(r) = \frac{c_0 r \ell_c^4}{(r^2 + \ell_c^2)^{5/2}}
        \phantom{A}_2F_1(\frac{5- \omega \ell_c}{2}, \frac{5+ \omega \ell_c}{2} ; 3 , \frac{\ell_c^2}{r^2 + \ell_c^2} )
\ee
with the corresponding quantization condition $\omega\ell_c = 5 + 2 k$. 
In this case also, the conditions at $r=0$ and $r=\infty$ are  obeyed.

No closed form solutions exist, to our knowledge, for the coupled system 
(i.e. for $\kappa > 0$), the exact values $\omega=4$ and $\omega=5$
are recovered in the limit of vanishing  scalar field; this limit is indeed 
equivalent to the limit $\kappa \to 0$ through an appropriate rescaling.

The metric approaches the form (\ref{ads}) asymptotically 
and the scalar field's decay is powerlike:
\be
         \phi(r)_{r \to \infty} = \frac{\langle\phi\rangle}{r^a} \ \ , \ \ a = 2 + \sqrt{4 + m^2}
\ee
where a mass term of the scalar field, say $m$, has been included for completeness and the 
notation $\langle\phi\rangle$ is used along many papers e.g. \cite{Dias:2011at}.

\subsection{Chern-Simons limit: $U=0$, $\alpha = \ell^2/2$}
As pointed out above, the solutions for $\alpha >0$, $\ell^2 < \infty$
exist only for $2 \alpha/\ell^2 < 1$. The limit $\alpha = \ell^2/2$
is special and known as the Chern-Simons (CS) limit \cite{chamseddine} (see also \cite{Brihaye:2013vsa}).
The asymptotic form of the metric in the CS limit is special. For generic values of $\alpha$, we have 
\be
                f(r) =  \frac{r^2}{\ell_c^2} + 1 + \frac{f_2}{r^2} + O(1/r^4) \ \ , 
                      \ \   b(r) =  \frac{r^2}{\ell_c^2} + 1 + \frac{b_2}{r^2} + O(1/r^4) \ \, \ \  
                      h(r) = r^2\left( 1 + \frac{h_2}{r^4}\right) \ \   
\ee
where $\ell_c$ is defined above. The mass  is given by 
\be  
 M = \frac{V_3}{16 \pi G} (f_2-4 b_2) \sqrt{1 - \frac{2\alpha}{\ell^2}} \ .
\label{mass_generic} 
\ee
which becomes undefined for $\alpha = \ell^2/2$. By contrast, in the Chern-Simons limit, we have \cite{Brihaye:2013vsa}~:
\be
                f(r) =  2{r^2} + 1+ {\cal V} + \frac{\tilde f_2}{r^2} + O(1/r^4) \ \ , 
                      \ \   b(r) =  2 {r^2} + 1 +{\cal U}+ \frac{\tilde b_2}{r^2} + O(1/r^4) \ \, \ \ 
                           h(r) = r^2\left( 1 + \frac{\tilde h_2}{r^2}\right)                       
\ee
The mass of the solution is obtained in terms of ${\cal U,V}$ (see \cite{Brihaye:2013vsa} for details). 
In the non spinning case,
we have ${\cal V} = {\cal U}$ and $M = \frac{ V_3}{8 \pi G} \frac{3({\cal U}^2-1)}{8}$;
our numerical results nicely confirmed that this value is approached  by (\ref{mass_generic}).

\section{Non-spinning Solutions}
As in the previous section discussed, the system of non-linear equations does not admit in general (to our knowledge) solutions in closed
form and some numerical techniques need to be employed to construct the solutions.
We used the routine COLSYS  \cite{colsys} based on the Newton-Raphson algorithm with adaptive grid scheme.

The effect of   the Gauss-Bonnet term on five   dimensional (non rotating) boson  stars was 
analyzed in \cite{Hartmann:2013tca}   with a $\phi^6$ potential.
Many  features of the solutions are qualitatively similar with the potential (\ref{susy_pot}). 
In particular, it turns out  that the numerical routine used to construct the solutions
becomes inefficient for the large values of $\phi(0)$
when  the parameters $\kappa$, $\alpha$ are   chosen in the same range of magnitude.
We attempted to understand the reason of this technical
difficulty by inspecting  the behaviour at the   origin of the metric and of the Ricci scalar $R(r)$.
In complement to the figures presented in \cite{Hartmann:2013tca}, we present  on $\varA{Fig. \ref{non_rot}}$
the values of $R(0)=-4(3f''(0)+b''(0)/b(0))$ and the frequency $\omega$ for $\kappa=0.1$ and three values of $\alpha$
(for instance $\alpha = 0.0; 0.1; 1.0$).  

We first discuss the solutions for Einstein gravity (i.e. for $\alpha=0$). 
For fixed $\kappa$ and increasing $\phi(0)$ a family of solutions can 
be constructed with the following  features~:
\begin{enumerate}
\item The value $b(0)$ rapidly approaches zero.
\item The values $-f''(0)$ and $-R(0)$ increases significantly. 
\item The frequency $\omega$ approaches a constant.
\end{enumerate}
In particular the Ricci scalar is negative for the full space-time . 

The strong variations of the metric fields $f(r)$, $b(r)$ around $r=0$ cause numerical instabilities.
As an example, setting  $\kappa = 0.1$ (e.g. see $\varA{Fig. \ref{non_rot}}$) 
we find $b(0) \sim 10^{-3}$, $f''(0) \sim - 10^4$, $R(0) \sim - 10^{5}$, 
$\omega \approx 0.89$ for $\phi(0) = 8$. For  $\phi(0) > 10 $ the numerical results
become unreliable.   
A configuration
presenting an essential singularity at the origin is likely approached 
(roughly speaking 'exponentially') when the central
value of the scalar field increases. Our results however cannot predict whether the
limiting configuration is reached for a finite or infinite value of $\phi(0)$.

However we believe that the set of solutions that we can construct with out technique
catch the essential part of the whole pattern.
Indeed, in the present case and as well as for  the spinning solitons (discussed below), 
the $\varA{Q-M}$ plot of the mass of the solutions versus the particle number $Q$  reveals that
the solutions assemble in two branches forming a spike at the maximal value of $Q$, say $Q = Q_M$ 
(see e.g. several Figs. in \cite{Hartmann:2012gw}). 
For a given value of the particle number $Q$, the solutions with the lowest mass lies on
the branch directly connected to the vacuum (corresponding to the small values of $\phi(0)$);
the solutions  corresponding to the high values of $\phi(0)$ have a slightly higher mass 
and, therefore, are likely  unstable.


\begin{figure}[h]
\begin{center}
{\label{non_rot_1}\includegraphics[width=8cm]{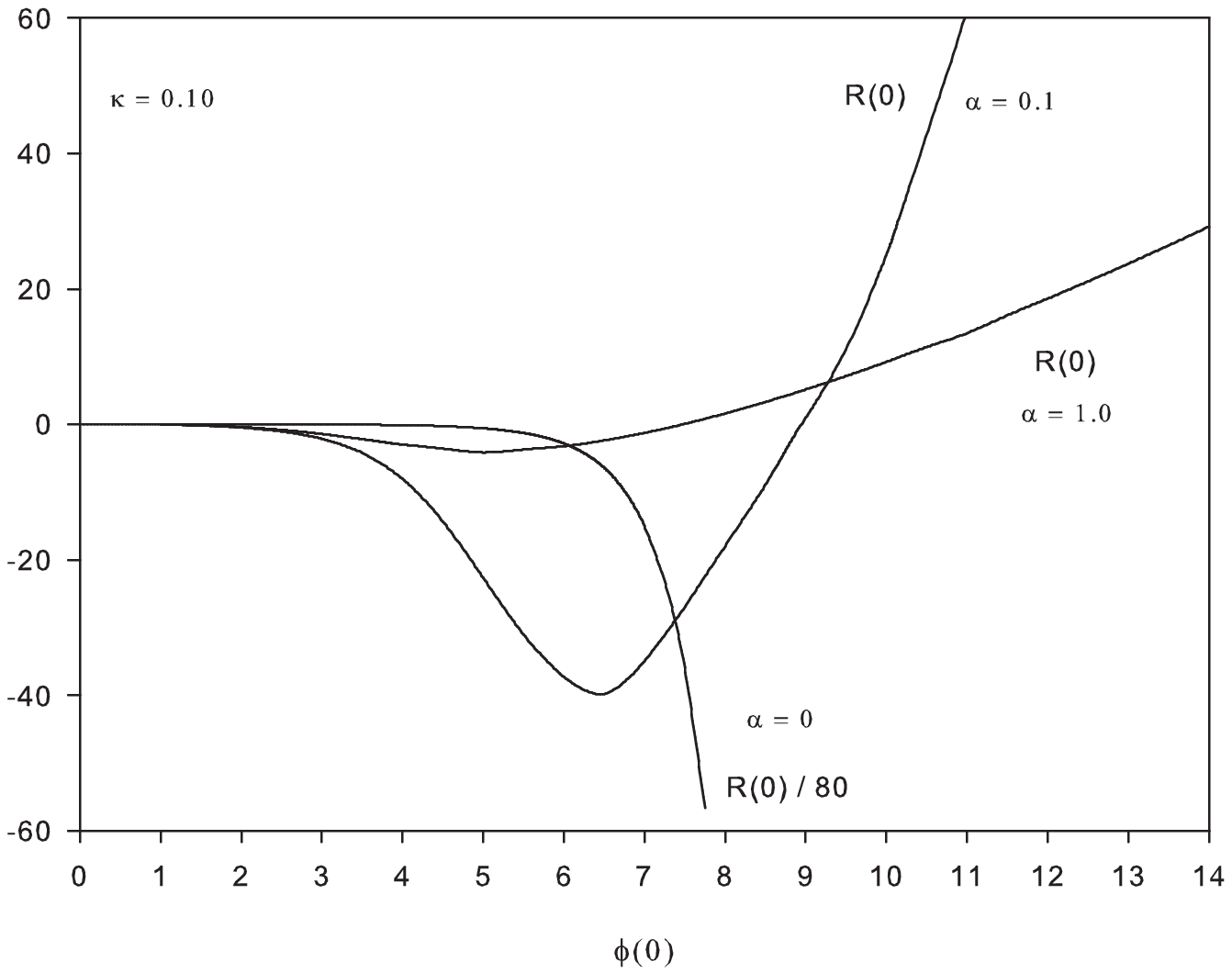}}
{\label{non_rot_2}\includegraphics[width=8cm]{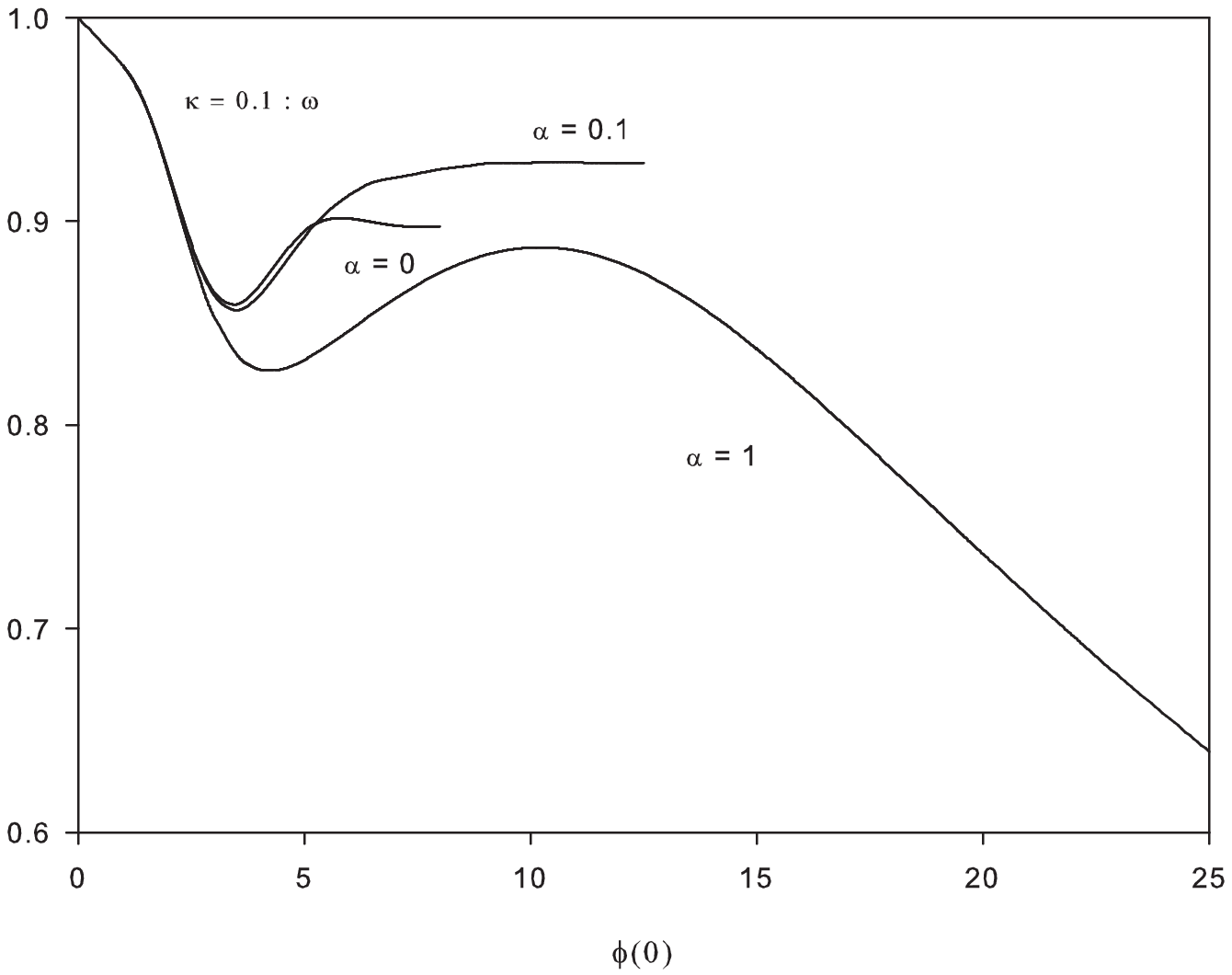}}
\end{center}
\caption{The Ricci scalar at the origin $R(0)$  (left) 
and the frequency (right) as function of   $\phi(0)$ 
for $\kappa=0.1$  and for three values of the Gauss-Bonnet coupling  $\alpha=0.0, 0.1,$ and $\alpha = 1.0$.
\label{non_rot}
}
\end{figure}

Increasing gradually the Gauss-Bonnet parameter reveals that the properties (1)  and (3) above 
 still  holds. By contrast, $R(0)$ 
 changes signs (becoming positive)  and  increases significantly for larger values of $\phi(0)$. 
 This is illustrated be the $\alpha=0.1$-line in $\varA{Fig. \ref{non_rot}}$.
One effect of the Gauss-Bonnet term 
  is then a change of the sign of the curvature of space-time in the core of heavy solitons:
  for $r < r_{core}$, we have
  $R(r) \leq  0$ for $\alpha=0$, $R(r) \geq 0$ for $\alpha >0$.
  
 Finally, for large values of $\kappa$ (typically setting $\alpha >  \kappa$), 
 the parameter $\omega$ keeps varying significantly with $\phi(0)$
  (as indicated by $\varA{Fig. \ref{non_rot}}$ for the case $\alpha = 1$)
 and the numerical integration is easier.  
 A singular configuration seems to be approached as well, 
 but the convergence is slower than for the small values of $\alpha$. 
 Setting $\alpha = 10$ and $\kappa =  1.0$, for example, robust solutions 
 can be  constructed up to $\phi(0) \sim 50$).

The behaviour of
the quantities $M,Q$  as functions of the frequency $\omega$ is largely discussed by
 the figures presented in \cite{Hartmann:2013tca}. It is however  useful to remember
 some features for the next sections. 
The $\varA{\omega-M}$ plot  reveals a series of branches assembling in the form of a spiral. 
For the low values of $\alpha$ (including  Einstein gravity $\alpha=0$) the
solitons exist for frequencies $\omega$ larger than a minimal value $\omega_m$ (this value off course
depends on  $\kappa$ and of $\alpha$). The quantities $M,Q$ are bounded.

Increasing gradually the Gauss-Bonnet parameter $\alpha$ has the effect to
'unwind' the spiral. When $\alpha$ is  chosen of the same order as $\kappa$, 
only one branch of solution survives extending back to the small values of $\omega$.

\section{Spinning Solutions: Case $\Lambda = 0$}
\subsection{Flat space}
To calibrate the  numerical results,
we first studied the solitons in the case $\kappa=0$, $\alpha=0${\bf ,}
i.e. the rotating Q-balls interacting through the SUSY
potential.
The results can then be compared to the ones of \cite{Hartmann:2010pm} 
(our  potential is bounded for $\phi \to \infty$ while the potential used in \cite{Hartmann:2010pm} 
grows like $\phi^6$). 
For spinning solutions, the relevant shooting parameter labeling the families of solutions is $\phi'(0)$.
As expected, our results are qualitatively  similar results to \cite{Hartmann:2010pm} 
for the small values of $\phi'(0)$; 
the corresponding scalar function is indeed small enough and  confined in the region where the
polynomial and SUSY potentials coincide.

On $\varA{Fig. \ref{gb_0}}$ we plot the mass, the  charge (left side) and the frequency $\omega$ (right side)
 as functions of the parameter $\phi'(0)$. 
In the limit $\phi'(0) \to 0$ the solution corresponds to the  thick wall limit  and 
it is not well localized around the origin
because $\omega \to 1$; at the same time   the mass and charge increase.  
 
 The parameter $\phi'(0)$  cannot be increased arbitrarily; indeed 
 no solutions can be constructed for $\phi'(0) > 2.8$. At the approach of this maximal value both the mass
 and the charge increase considerably and likely tend to infinity.  
 The numerical results reveal that the increase of $M$ and $Q$ is related to the fact that both the range and the
 amplitude of the scalar field increase although the solution is exponentially localized. When the maximal value
 of $\phi'(0)$ is approached,   the frequency $\omega$ reaches a plateau at $\omega \approx 0.0697$.
 The situation is nevertheless different from the case of non-spinning solutions (see previous section)
where the mass and the charge are bounded while $\omega$ is constant. 
 
 To finish, we  note that
 the frequency $\omega$  decreases monotonically while $\phi'(0)$ increases; this  contrasts 
   with case of a the polynomial potential (see e.g. $\varA{Fig. 1}$ of \cite{Hartmann:2010pm}).
   This different behavior can be explained: with a  bounded potential, the 'energy cost' for the scalar field
   to be large in some domain of space is smaller than for a polynomial potential.

\begin{figure}[h]
\begin{center}
{\label{gb_0_2}\includegraphics[width=8cm]{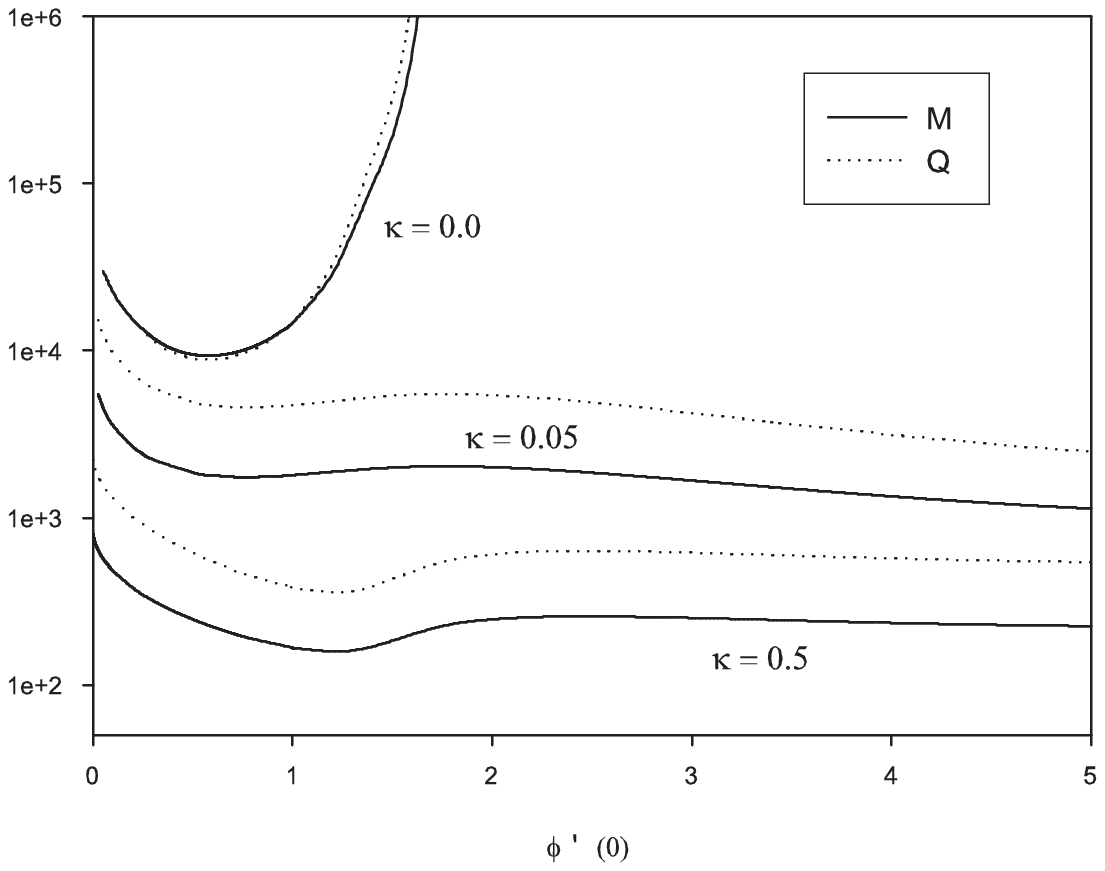}}
{\label{gb_0_3}\includegraphics[width=8cm]{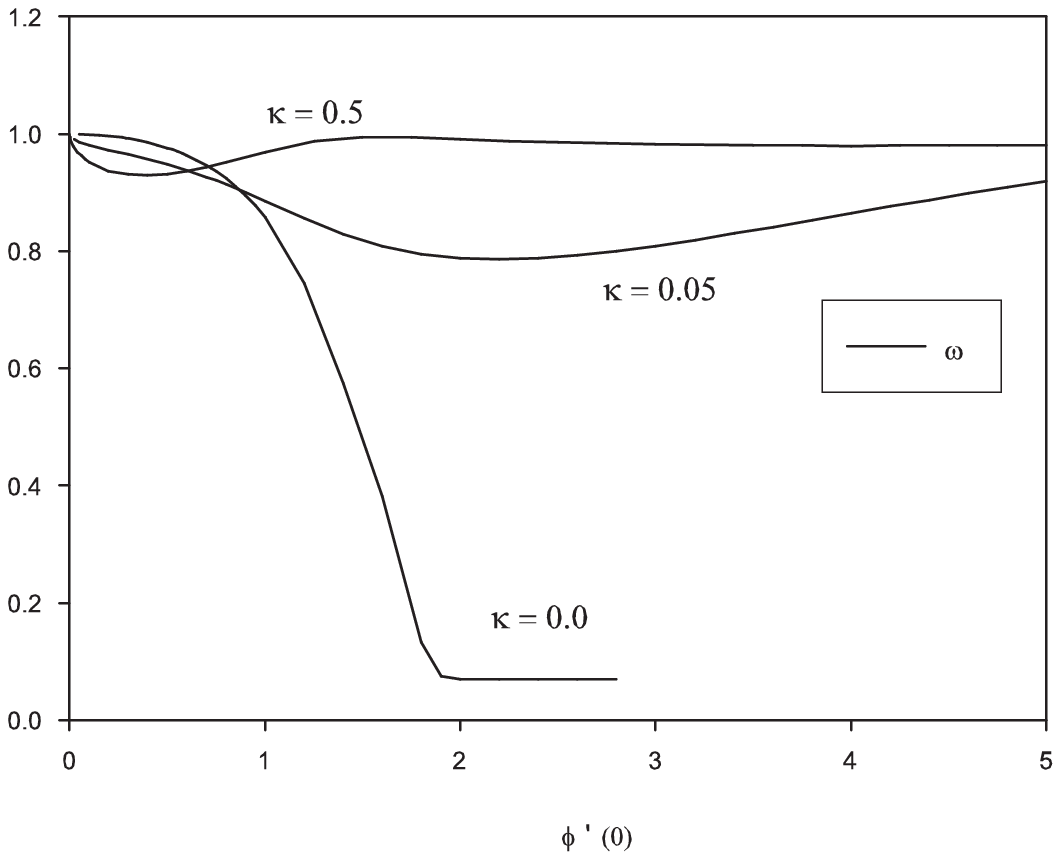}}
\end{center}
\caption{Mass, Charge  (left) 
and the frequency (right) as function of $\phi'(0)$ 
for three values of Newton coupling $\kappa$ and for $\alpha=0.0$.
\label{gb_0}
}
\end{figure}

Because the effect of the Gauss-Bonnet parameter is especially apparent when the
mass and charge of the soliton are plotted in function of the frequency $\omega$,
we find  it useful to supplement such a plot, see $\varA{Fig. \ref{gb_0_1}}$.

\begin{figure}[h]
\centering
\epsfysize=10cm
\mbox{\epsffile{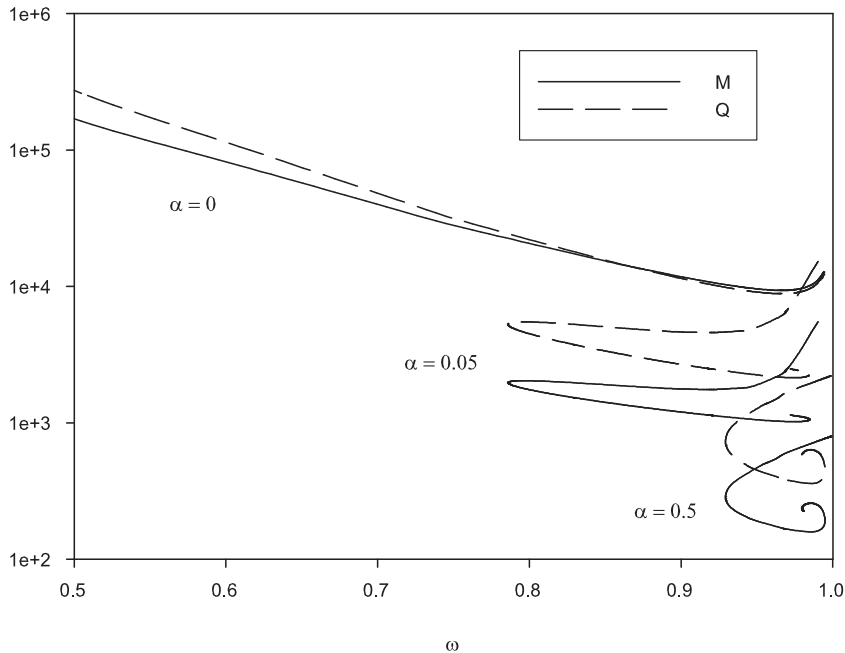}}
\caption{\label{gb_0_1}
Mass and charge of the rotating Q-balls as function of the frequency
for three values of Newton coupling $\kappa$ and for $\alpha=0.0$.}
\end{figure}

\subsection{Einstein gravity}
Setting $\kappa > 0$, the energy momentum tensor of the soliton
constitutes a source term for gravity, the solutions are then boson stars. 
In this paper, we studied in 
particular the cases corresponding to  $\kappa = 0.05$ and $\kappa = 0.5$ with the hope that
it catches the main pattern.
As pointed out in \cite{Hartmann:2010pm} the pattern
of the gravitating solutions  is qualitatively  different from the ones living in flat space-time. 
One of the main difference   (demonstrated by $\varA{Fig. \ref{gb_0}}$)
is that the gravitating solutions can be constructed for larger values
of the parameter $\phi'(0)$. When the parameter $\phi'(0)$ becomes too large,
the numerical integration is problematic because the value $b(0)$ becomes very small
(we could reach up to $b'(0) \sim 0.0001$ without problem).
At the same time, the value  of the Ricci scalar at the origin $R(0)$ and $|f''(0)|$ increases
regularly with $\phi'(0)$ as indicated by $\varA{Fig. \ref{critical_rot}}$.
\begin{figure}[h]
\begin{center}
\includegraphics[width=10cm,angle=0]{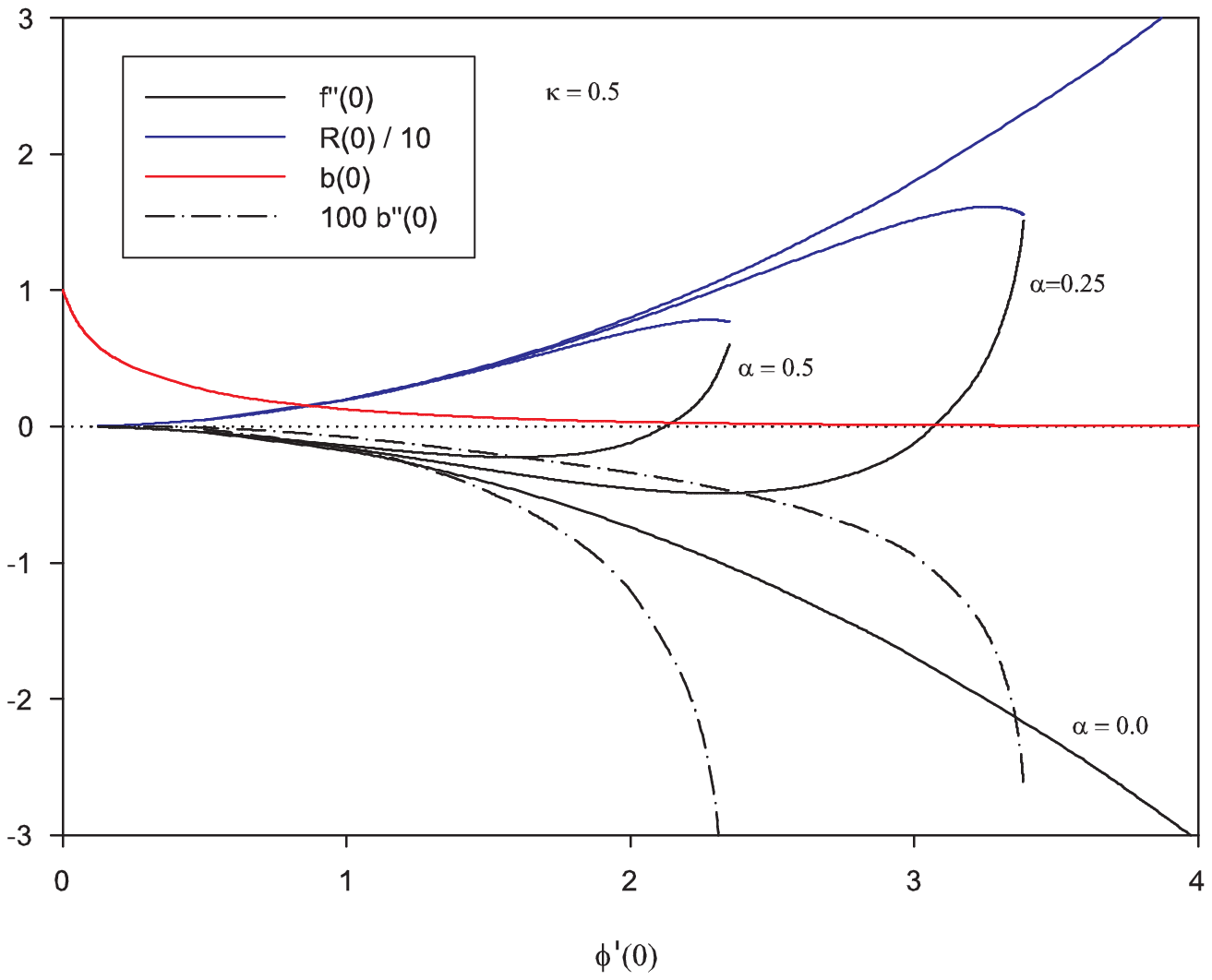}
\end{center}
\caption{\label{critical_rot} Several quantities characterizing the metric at the origin 
as function of $\phi'(0)$
for  $\kappa=0.5$ and $\alpha=0.0, 0.25,$ and $\alpha =  0.5$.}
\end{figure}
These numerical observations suggest  that a
configuration presenting  a naked singularity at the origin is approached in the limit $\phi'(0) \to \infty$. 
We could not find an algebraic argument  demonstrating this statement. 

When the mass and the charge are  plotted as function of the frequency $\omega$, the graph reveals the occurence of
 several branches of solutions existing on different intervals of  the frequency $\omega$ 
 and forming a spiral.   
The first branch (say $branch_1$) is connected to the vacuum (i.e. $\phi(r)=0, \omega=1$) exist for $\omega_1 \leq \omega \leq 1$.
A second branch exists (say $branch_2$) for  $\omega_1 \leq \omega \leq \omega_2$ and coincides with $branch_1$ at $\omega= \omega_1$.
Then a third branch exists for    {\bf $\omega_2 \leq \omega \leq \omega_3$}.  In fact several secondary branches likely  exist 
but are not displayed on the figure. Setting, for example, $\kappa = 0.05$, we find $\omega_1 \approx 0.785$,  $\omega_2 \approx 0.98$, and $\omega_3 \approx 0.94$.
In the interval of frequencies where two or more branches coexist the solutions of $branch_2$ are  the ones  with the lowest mass; the interval $[\omega_1, \omega_2]$
therefore plays a fundamental role, since it   'hosts' the solutions with the lowest energy. 

%
%

\subsection{Einstein-Gauss-Bonnet gravity}
We now discuss the effect of the Gauss-Bonnet term on the spinning soliton.
The results reveal that, for spinning boson  stars, 
 the deformation of the Einstein soliton by the Gauss-Bonnet term 
obeys a  different pattern with respect to the non spinning case. 
This effect is illustrated  by $\varA{Fig. \ref{mass_over_omega_005}}$
where $\kappa = 0.05$ is  choosen for definiteness. 

\begin{figure}[h]
\begin{center}
\includegraphics[width=8cm,angle=270]{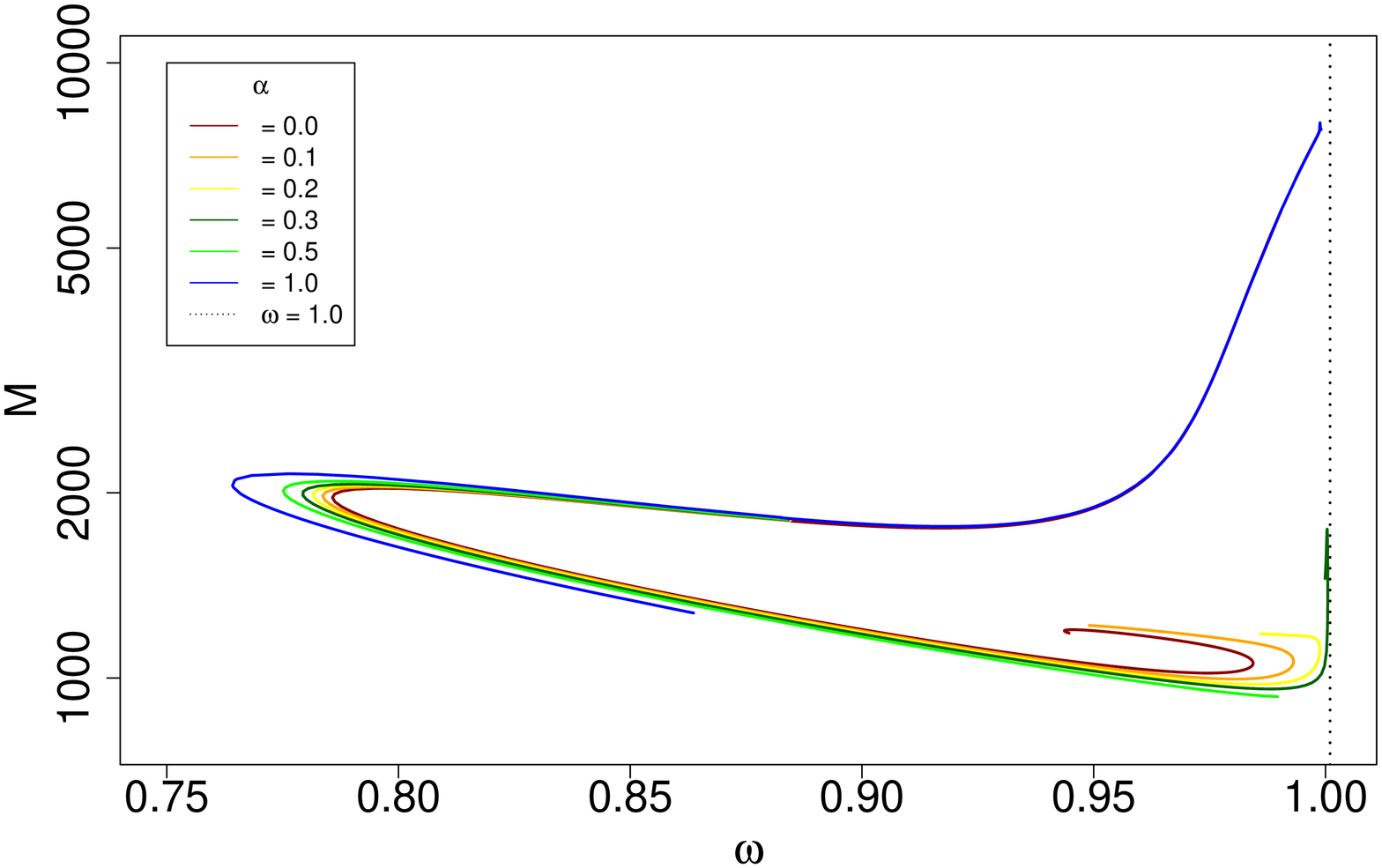}
\end{center}
\label{mass_over_omega_005}
\caption{\label{mass_over_omega_005} Mass of rotating boson stars as function of the frequency
for  $\kappa=0.05$ and $\alpha=0.0$ (red)$, 0.1$ (orange)$, 0.2$ (yellow)$, 0.3$ (dark green)$, 0.5$ (green)$, $ and $\alpha = 1.0$ (blue).}
\end{figure}
\begin{figure}[h]
\begin{center}
\includegraphics[width=8cm,angle=270]{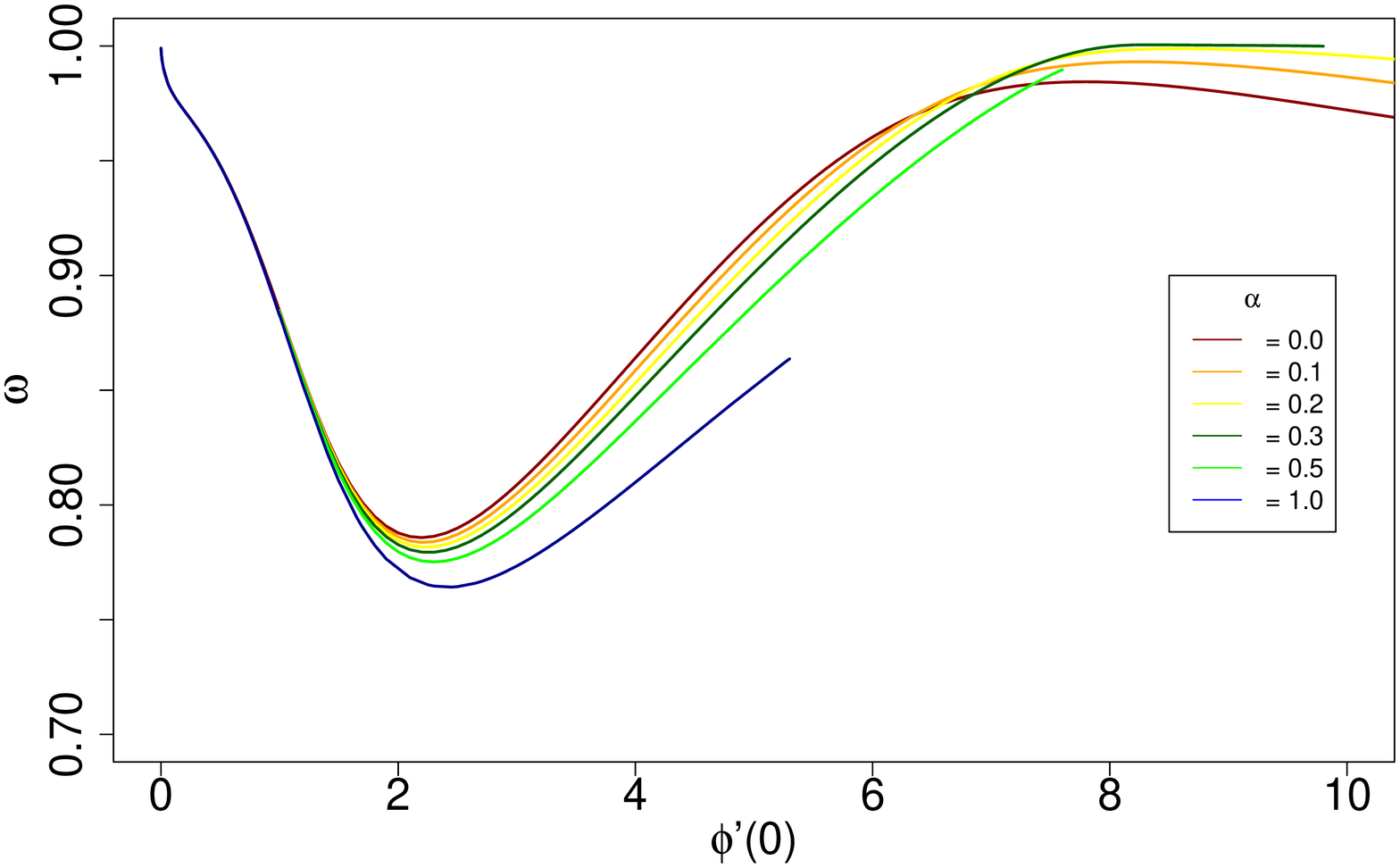}
\end{center}
\caption{\label{omega_over_phidash_005} The frequency of rotating boson stars as function of $\phi'(0)$
for  $\kappa=0.05$ and $\alpha=0.0$ (red)$, 0.1$ (orange)$, 0.2$ (yellow)$, 0.3$ (dark green)$, 0.5$ (green)$, $ and $\alpha = 1.0$ (blue).}
\end{figure}

Using the notations defined in the previous section, we focus mainly on the
fundamental interval $[\omega_1,\omega_2]$. It  turns out that, increasing $\alpha$, the lower limit $\omega_1$
of the spiral decreases while the upper limit $\omega_2$
increases. The 'range of frequencies' of solutions with minimal energy therefore increases with $\alpha$.
In fact the upper limit of this main branch reaches and slightly overtakes the critical value $\omega=1$. 
This is not a contradiction with the limit $\omega=1$ (see (\ref{bessel_2})) 
  because the corresponding solution is strongly gravitating.
Such a phenomenon has, to our knowledge, not been observed so far in the study of boson stars. 
The numerical results show that the above pattern holds for generic values of $\kappa$.
However, when $\kappa$ gets large enough (typically $\kappa=\alpha$)
the spiral remains far enough from the value $\omega=1$.

As for the non-spinning case, the construction of the solutions in the large $\phi'(0)$ limit 
is technically challenging.  
The structure of  solutions reached while increasing $\phi'(0)$ appears  
quite different from the $\alpha=0$ case.
One aspect of this  phenomenon  is  illustrated namely  
 on $\varA{Fig. \ref{omega_over_phidash_005}}$ where 
the dependance of the frequency over the parameter $\phi'(0)$ is presented. 
It shows that the curves apparently terminates at some maximal values of $\phi'(0)$ (depending on $\alpha$)
where, once more,  the numerical integration becomes very problematic.
A natural explanation of this phenomenon is provided by $\varA{Fig. \ref{critical_rot}}$.
Here the second derivatives $f''(0),b''(0)$ and the value $R(0)$ of the Ricci scalar at the origin
are reported as functions of $\phi'(0)$ for $\kappa = 0.5$ and several values of $\alpha$. 
It strongly suggests that when $\phi'(0)$ approaches a 'numerical' maximal value,
the second derivatives  increase  significantly and even seem to become infinite for some maximal value, say $\phi'(0)=K_c(\kappa,\alpha)$. 
At the same time the value $R(0)$ remains  finite, suggesting that the limiting configuration is regular. 

As a conclusion, our results indicate that (i)
 spinning solitons in EGB gravity  exists up to a maximal value
say $\phi'(0) = K_c(\kappa,\alpha)$ (which depends off course on $\kappa,\alpha$);
 (ii) they present  a maximal value of the mass, the charge
and the angular momentum; (iii) the limiting configuration is regular and the  divergence of $b''(0)$ 
and $f''(0)$ reflect an  singularity due to the gauge fixing of the radial coordinate. 

\clearpage
\section{Boson stars for $\Lambda < 0$, $U=0$}
In the absence of a potential, one can take advantage of the rescaling of the radial variable $r$
and of the scalar field to set $\ell$ and $\kappa$ to particular values without loosing generality.
Here, we have choosen $\ell = 1$ and $\kappa = 0.05$. The results presented in this section are obtained mainly 
for $U=0$; the solutions with the potential (\ref{susy_pot}) will be presented in \cite{riedel}.
\begin{figure}[h]
\begin{center}
{\label{non_rot_1}\includegraphics[width=8cm]{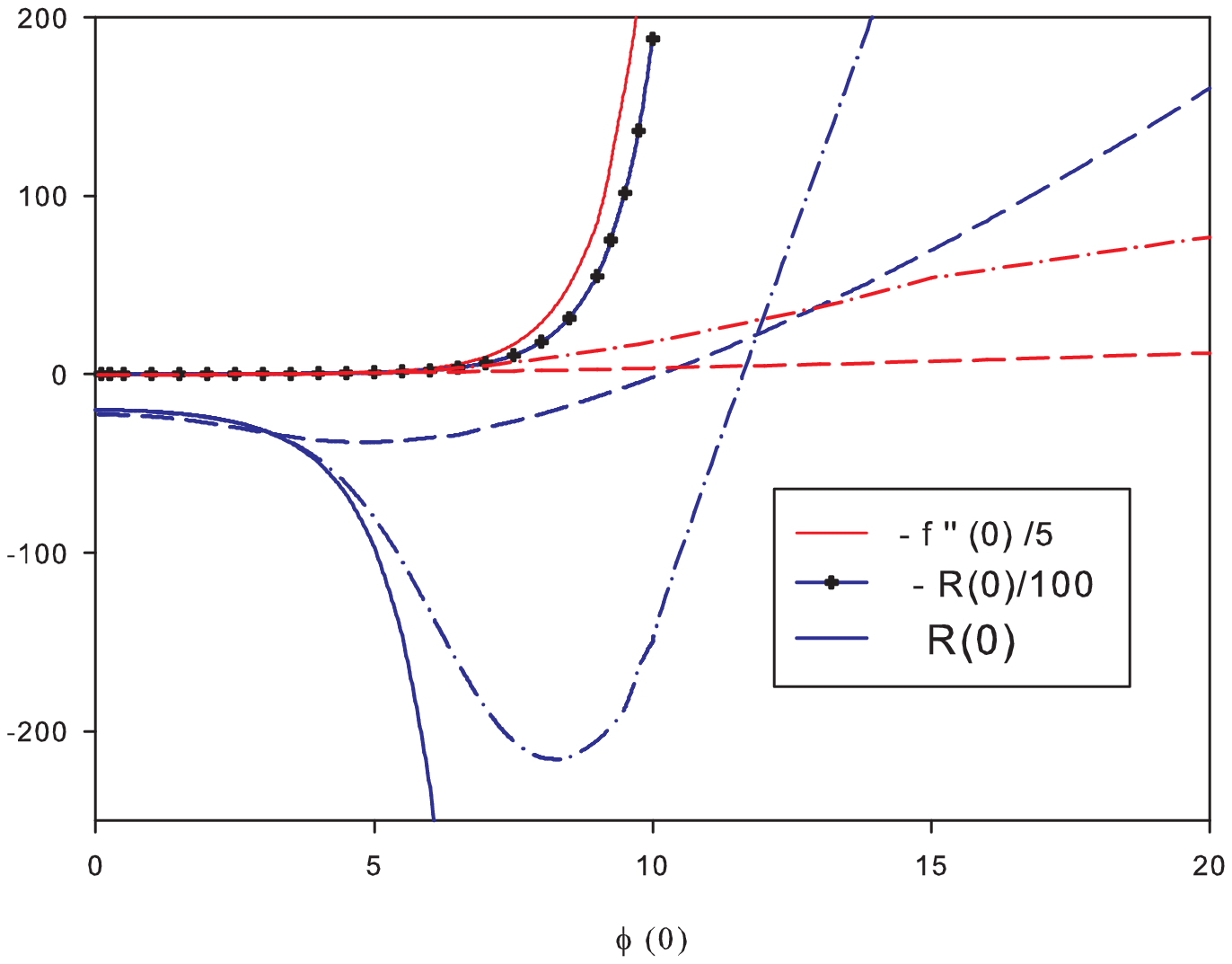}}
{\label{non_rot_2}\includegraphics[width=8cm]{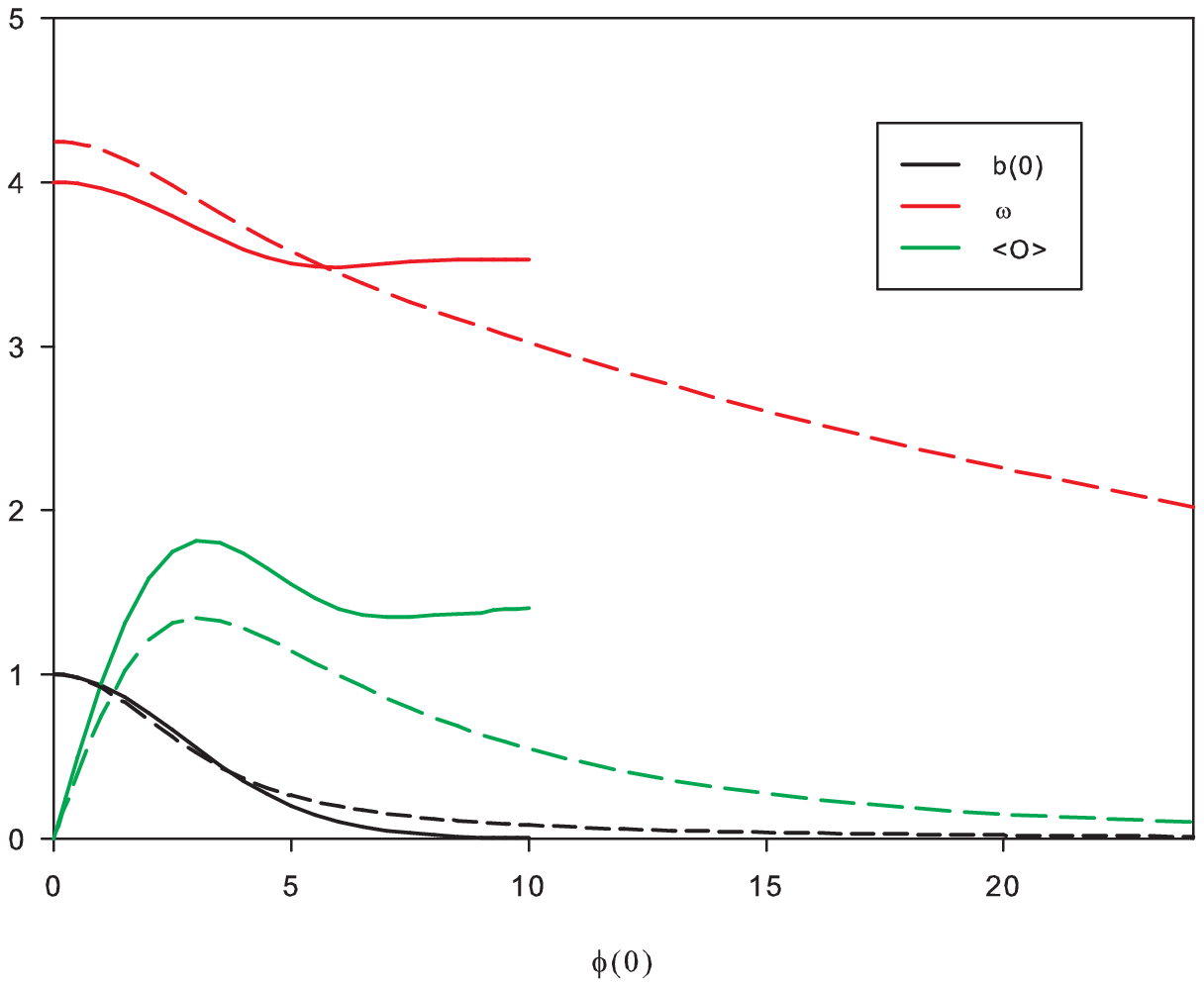}}
\end{center}
\caption{Left: The values $R(0)$ and $f''(0)$ versus  $\phi(0)$ for $\kappa=0.05$
and $\alpha = 0$ (solid lines), $\alpha = 0.02$ (dot-dashed lines) and $\alpha = 0.2$ (dashed lines).
Right:  b(0), the frequency $\omega$, and the condensate $\langle O\rangle$ as function of   $\phi(0)$ 
for $\alpha = 0$ (solid lines) and $\alpha = 0.2$ (dashed lines).
\label{ads_1}
}
\end{figure}
We first discuss the non-spinning solutions. The pattern is  similar to the case of asymptotically flat solutions.

As noticed already, the Gauss-Bonnet parameter is bounded:~$0 \leq \alpha \leq \ell^2/2$.
Fixing $\alpha$ in this interval, 
 a family of solutions can be constructed and labelled by $\phi(0)$.
On $\varA{Fig. \ref{ads_1}}$,  several
  quantities characterizing the ADS solutions for pure Einstein gravity are presented by the solid lines
 and the effect of the Gauss-Bonnet term appears through the
 dashed-dotted lines (for $\alpha = 0.02$) and the dashed lines (for $\alpha = 0.2$).

The general features of the system pointed out in the asymptotically flat case (see previous section)
hold for AdS solutions. For $\alpha=0$,  increasing the parameter $\phi(0)$ leads to configurations where
   the function $f(r)$ and the Ricci scalar $R(r)$ becomes very peaked at the origin.
   In particular the values $f''(0)$ and  $R(0)$ are negative and  decrease  
 considerably as shown by $\varA{Fig. \ref{ads_1}}$.  At the same time, the value $b(0)$ becomes quite small. 
Setting for definiteness $\kappa = 0.05$, we could construct 
reliable solutions for $\phi(0) < 10$ as reflected on both sides of $\varA{Fig. \ref{ads_1}}$:  
   on the left side, the line with bullets demonstrate that both $R(0)$ and $f''(0)$ diverge simultaneously, 
   on the right side the solid curves stop. 
Complementing $\varA{Fig. \ref{ads_1}}$, the dependance of $M$ and $Q$ on the frequency $\omega$ is shown 
by the solid lines of the left side of $\varA{Fig. \ref{ads_2}}$.    

The requirement of the regularity conditions at the origin, together with the ADS asymptotic 
conditions makes it   difficult to obtain reliable  results for large values of $\phi(0)$ but  
we think however that the part of the  branch  that we obtained reflects the general pattern. 
 
  For $\alpha > 0$, the sign of the Ricci scalar in the core of the soliton  becomes positive
  for high enough values of $\phi(0)$, i.e. when the soliton becomes heavier.  
  A singular-configuration seems to be approached as well but
  the approach is  softer  as demonstrated by $\varA{Fig. \ref{ads_1}}$.  The mass and the charge of the CS solitons
   have been supplemented on $\varA{Fig. \ref{ads_2}}$ (left side) by the dashed lines. The solutions exist for
   $\omega \leq 4 \sqrt{2} \approx 5.657$. 


The main features of the solutions persist in the presence of a potential.
The influence of the potential on the non-spinning solutions is sketched on $\varA{Fig. \ref{ads_2}}$ 
(right side) where
the solid lines refer to the solution with no potential and the dashed ones to the 
potential (\ref{susy_pot}) with  $m=1$.

\begin{figure}[h]
\begin{center}
{\label{non_rot_1}\includegraphics[width=8cm]{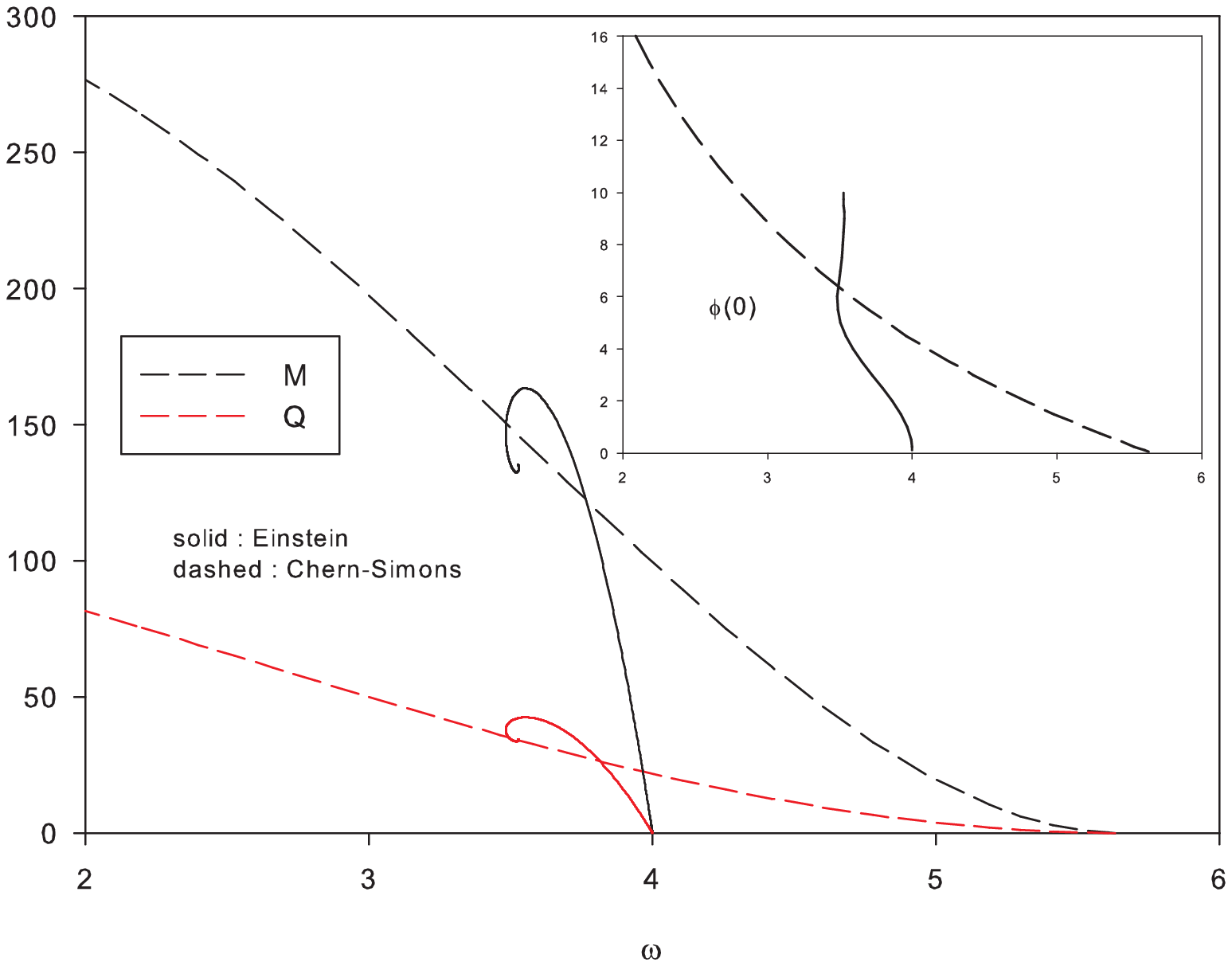}}
{\label{non_rot_2}\includegraphics[width=8cm]{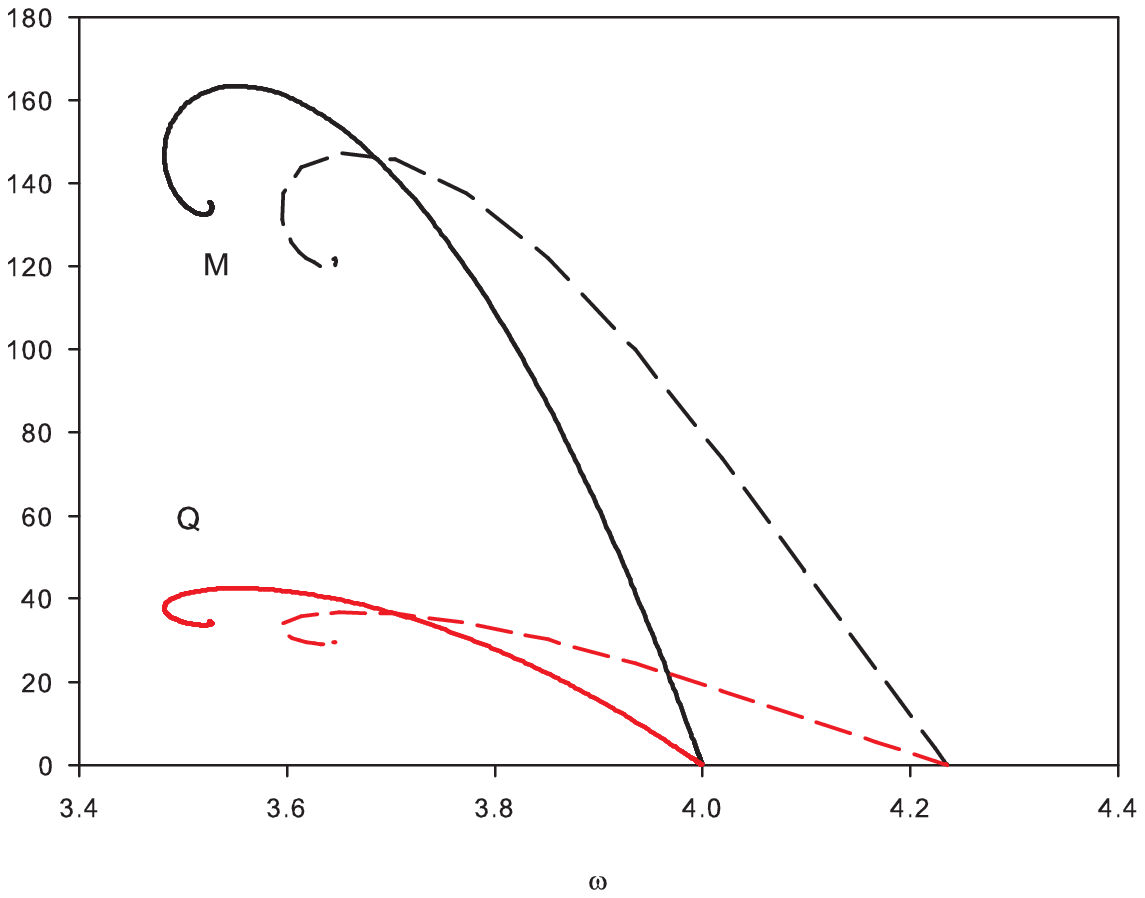}}
\end{center}
\caption{Left: 
The mass and the charge dependency on $\omega$ for the Einstein and Chern-Simons solitons
(solid and dashed lines respectively). The insert contains the corresponding value $\phi(0)$.
Right: The mass (black solid line) and charge (red solid line) versus $\omega$ for the family of Einstein solutions
of the left side figure (i.e. with no potential) and, in the dashed lines, the same quantities 
for the solutions with the potential 
(\ref{susy_pot}) with $m=1$. 
\label{ads_2}
}
\end{figure}
\begin{figure}[h]
\begin{center}
{\label{non_rot_1}\includegraphics[width=8cm]{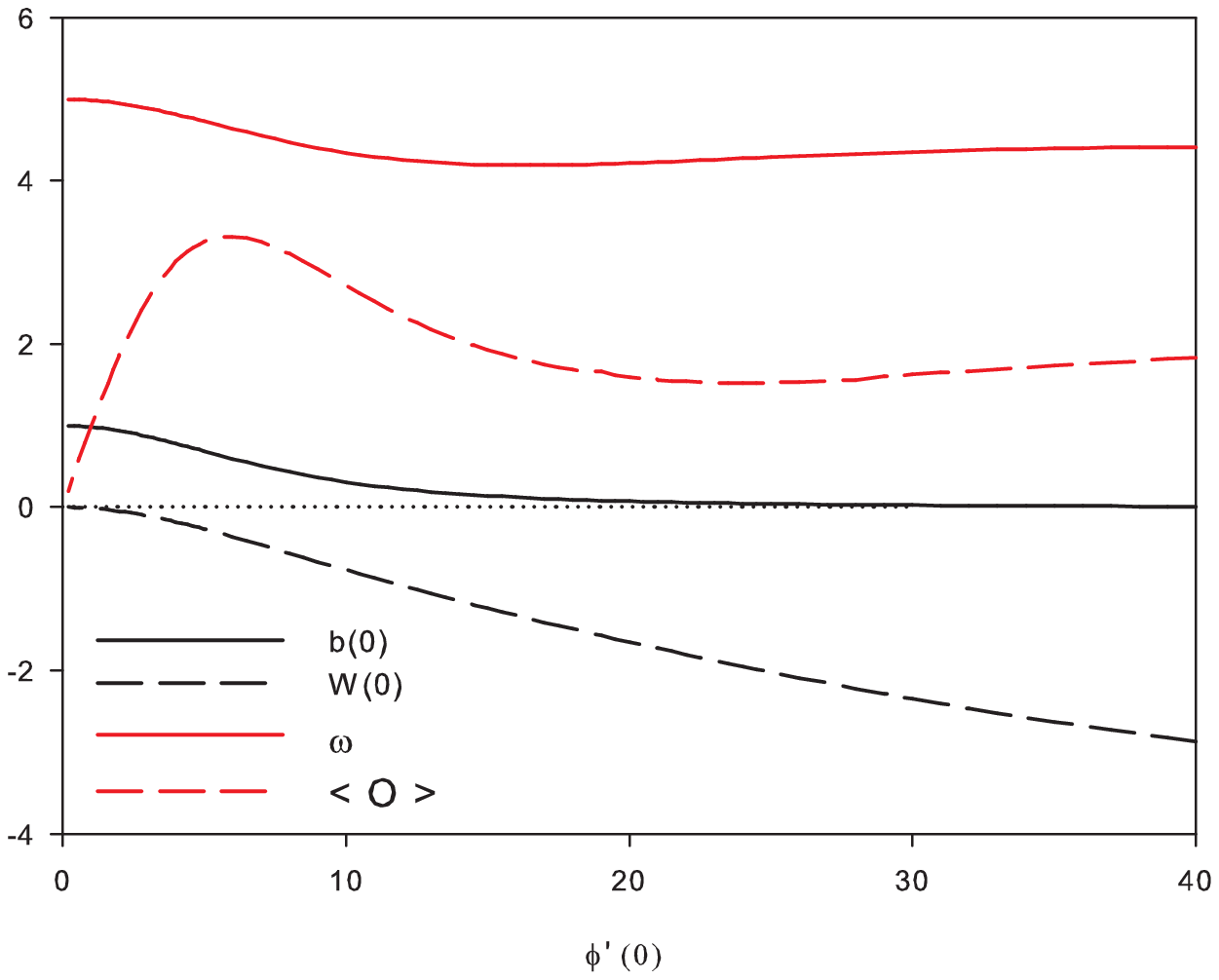}}
{\label{non_rot_2}\includegraphics[width=8cm]{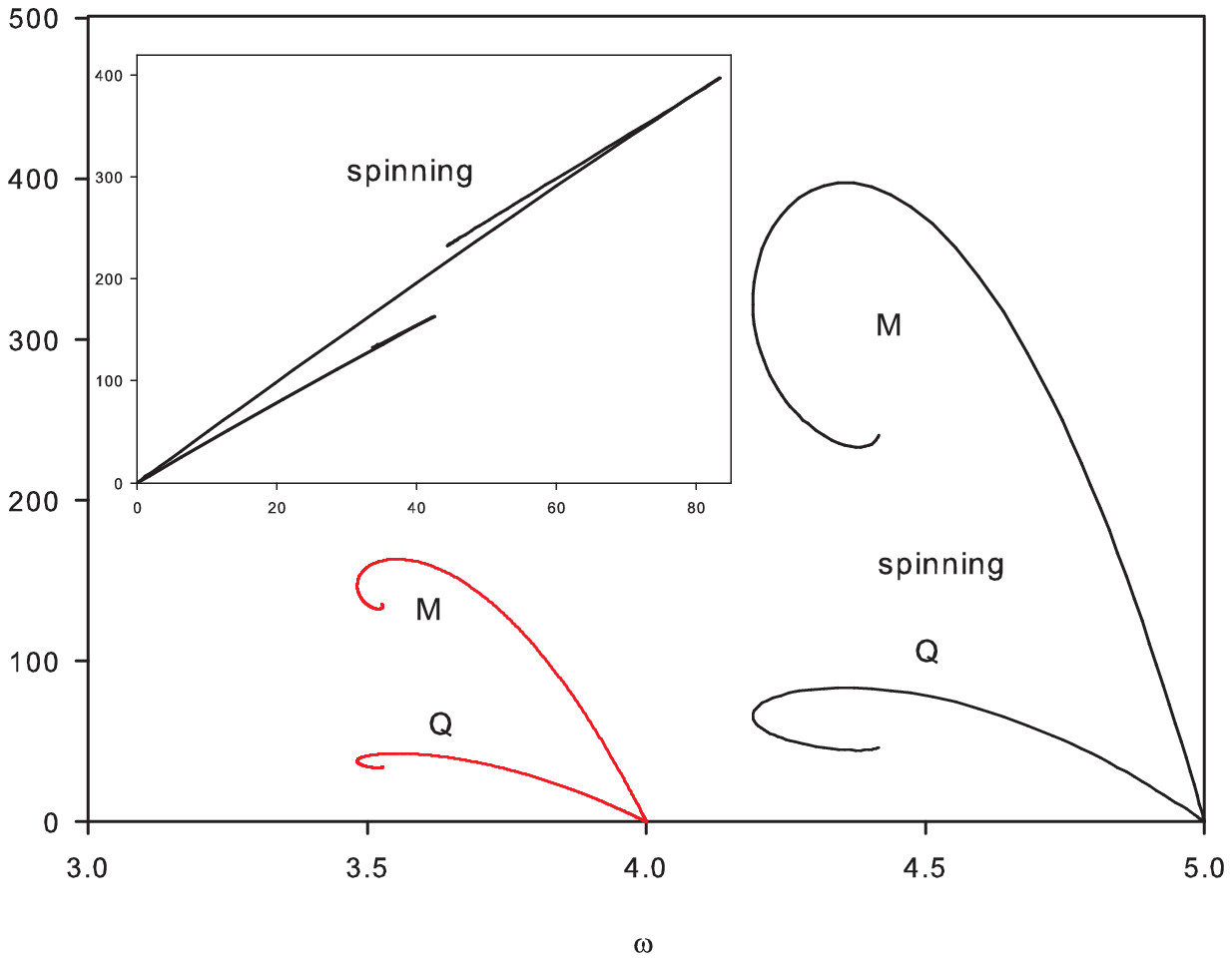}}
\end{center}
\caption{Left: The frequency $\omega$, the condensate $\langle O\rangle$, the value $W(0)$ and $b(0)$ as function of $\phi'(0)$ 
for $\kappa=0.05$ and $\alpha = 0$ for the spinning solutions with $m=0$.
Right: The mass (black line) and charge (red line) versus $\omega$ for the same values for spinning and non-spinning
solutions.
\label{ads_3}
}
\end{figure}

Finally we constructed AdS-spinning boson stars as well (for $U=0$). 
They present roughly the same features than their asymptotically flat counterparts
and, as expected, the same numerical difficulties occur. 

Several values characterizing the spinning boson stars corresponding to $\alpha=0$ are shown
on $\varA{Fig. \ref{ads_3}}$, left part. The value of the rotating function at the origin $W(0)$ 
becomes more negative while $\phi'(0)$ increases.
The dependance of the mass and charge for the non-spinning and spinning 
solutions for $\alpha =0$ is presented on the right side of the figure; the two solutions
exist on disjoined interval of $\omega$ terminating at the two analytical values $\omega=4$
and $\omega=5$ respectively (note: the curve for the spinning case is presented in \cite{Dias:2011at}
up to a rescaling of $M$ and $Q$ due to our choice $\kappa = 0.05$).
The $\varA{Q-M}$
plot of this data,  supplemented in the insert, shows that the solutions corresponding to the large
values of $\phi'(0)$  are likely instable.

\section{Conclusion}
Once  considered in five dimensional Minkowski or AdS space-time, the Klein Gordon equation
of a doublet of massive scalar fields admits, depending on some ansatz for
the angular dependency, at least two analytic solutions which are naturally localized in space.
It is  natural to study the deformation of these
solitons when adding a suitable self-interacting potential and/or coupling the underlying
scalar field to gravity. The two solutions are then promoted to 
 stationary boson stars with or without an angular momentum. 
Here, we choose the Einstein-Gauss-Bonnet action for the gravitational part and 
 a potential inspired by supersymmetric extensions of the Standard Model for the self-interacting part.    

The results presented in \cite{Hartmann:2013tca} have been extended
to the case of rotating boson  stars. 
When fixing the coupling constants $\kappa, \alpha$, a family of solutions exist.
It can be labeled  by the value of the scalar field  at the origin in the case of non-spinning solutions
and by the {\it derivative} of the scalar field  at the origin in the case of spinning solutions.
In both cases, the construction of the solutions is technically difficult when this value
becomes large enough. Our numerical results reveal 
new types of limiting effects which are peculiar  to the Gauss-Bonnet interaction. 

Once classified with respect to the frequency parameter $\omega$, 
multiple families of solitons (spinning or not)  exist
on particular intervals of $\omega$. 
It seems to be a generic property that, on a mass-frequency plot, these branches
assemble in a spiralling shaped curve. 

Increasing the value of the  parameter $\phi(0)$ (for non-spinning)
or of $\phi'(0)$ (for spinning) results is a progression {\it inside} the spiral.
In the case of non rotating solitons, the Gauss-Bonnet interaction has a tendency
to  unwind the spiral and  lead, for sufficiently large values of $\alpha$
to a unique branch: one soliton for each $\omega$. For the rotating soliton
the effect of the Gauss-Bonnet term is rather to enlarge the domain of $\omega$
where the first few branches of the spiral exist.
The coupling to Gauss-Bonnet coupling constant  leads to families of solitons existing up to a maximal value
of $\phi'(0)$, limiting the  number of branches.
We obtained convincing arguments that the limiting solution is regular,
although some derivatives of the metric functions tend to infinity due to coordinate  artefacts.
\\ \\
{\bf Acknowledgments} We gratefully acknowledge discussions with B. Hartmann. J.R. 
gratefully acknowledges support within the framework of the DFG Research
Training Group 1620 {\it Models of gravity}. 


\clearpage
\begin{thebibliography}{30}
\bibitem{fls} R. Friedberg, T. D. Lee and A. Sirlin, Phys. Rev. D {\bf 13} (1976) 2739.
\bibitem{lp} T. D. Lee and Y. Pang, Phys. Rep. {\bf 221} (1992), 251.
\bibitem{coleman}  S. R. Coleman, Nucl. Phys. B {\bf 262} (1985), 263.
\bibitem{vw} M.S. Volkov and E. W\"ohnert, Phys. Rev. D {\bf 66} (2002), 085003.
\bibitem{kk1} B. Kleihaus, J. Kunz and M. List, Phys. Rev. D {\bf 72} (2005), 064002.
\bibitem{kk2} B. Kleihaus, J. Kunz, M. List and I. Schaffer, Phys. Rev. D {\bf 77} (2008), 064025.
\bibitem{kusenko} A. Kusenko, Phys. Lett. B {\bf 404} (1997), 285; Phys. Lett. B {\bf 405} (1997), 108.
\bibitem{dm} {\it see e.g.} A. Kusenko, hep-ph/0009089.
\bibitem{ct} E. Copeland and M. Tsumagari, Phys.Rev. D {\bf 80} 025016 (2009).
\bibitem{cr} L. Campanelli and M. Ruggieri, Phys. Rev. D {\bf 77} (2008), 043504;
L. Campanelli and M. Ruggieri, Phys.~Rev.~D {\bf 80} (2009) 036006.
\bibitem{implications} K.~Enqvist and J.~McDonald, Phys.\ Lett.\ B {\bf 425} (1998), 309;
S.~Kasuya and M.~Kawasaki, Phys.\ Rev.\  D {\bf 61} (2000), 041301;
A.~Kusenko and P.~J.~Steinhardt, Phys.\ Rev.\ Lett.\  {\bf 87} (2001), 141301;
T.~Multamaki and I.~Vilja, Phys.\ Lett.\  B {\bf 535} (2002), 170;
M.~Fujii and K.~Hamaguchi, Phys.\ Lett.\  B {\bf 525} (2002), 143;
M.~Postma, Phys.\ Rev.\  D {\bf 65} (2002), 085035;
K.~Enqvist, {\it et al.}, Phys.\ Lett.\  B {\bf 526} (2002), 9;
M.~Kawasaki, F.~Takahashi and M.~Yamaguchi, Phys.\ Rev.\  D {\bf 66} (2002), 043516;
A.~Kusenko, L.~Loveridge and M.~Shaposhnikov,
Phys.\ Rev.\  D {\bf 72} (2005), 025015;
Y.~Takenaga {\it et al.}  [Super-Kamiokande Collaboration],
Phys.\ Lett.\  B {\bf 647} (2007), 18; S.~Kasuya and F.~Takahashi, JCAP {\bf 11} (2007), 019.
\bibitem{Astefanesei:2003qy}
  D.~Astefanesei and E.~Radu,
  Nucl.\ Phys.\ B {\bf 665} (2003) 594
  [gr-qc/0309131].
\bibitem{Prikas:2004fx}
  A.~Prikas,
  Phys.\ Rev.\ D {\bf 69} (2004) 125008
  [hep-th/0404037].
  \bibitem{Hartmann:2012gw}
    B.~Hartmann and J.~Riedel,
    Phys.\ Rev.\ D {\bf 87} (2013) 4,  044003
    [arXiv:1210.0096 [hep-th]].
  
  \bibitem{Hartmann:2010pm}
    B.~Hartmann, B.~Kleihaus, J.~Kunz and M.~List,
    Phys.\ Rev.\ D {\bf 82} (2010) 084022
    [arXiv:1008.3137 [gr-qc]].
  
\bibitem{Dias:2011at}
  O.~J.~C.~Dias, G.~T.~Horowitz and J.~E.~Santos,
  JHEP {\bf 1107} (2011) 115
  [arXiv:1105.4167 [hep-th]].
  
\bibitem{chamseddine}
A. H. Chamseddine, Phys. Lett. {\bf B233} (1989) 291.  
  
\bibitem{Brihaye:2013vsa}
  Y.~Brihaye and E.~Radu,
  arXiv:1305.3531 [gr-qc], JHEP {\bf 1311} (2013) 049.
  
  
\bibitem{Stotyn:2013yka}
  S.~Stotyn, C.~D.~Leonard, M.~Oltean, L.~J.~Henderson and R.~B.~Mann,
  ``Numerical Boson Stars with a Single Killing Vector I: the $D\ge5$ Case,''
  arXiv:1307.8159 [hep-th].
  
  \bibitem{Hartmann:2013tca}
    B.~Hartmann, J.~Riedel and R.~Suciu,
    arXiv:1308.3391 [gr-qc].
\bibitem{kunz1}  J.~Kunz, F.~Navarro-Lerida and A.~K.~Petersen,
  Phys.\ Lett.\ B {\bf 614} (2005) 104
  [gr-qc/0503010].  

\bibitem{Brihaye:2008kh}
  Y.~Brihaye and E.~Radu,
  Phys.\ Lett.\ B {\bf 661} (2008) 167
  [arXiv:0801.1021 [hep-th]].
  
\bibitem{Copeland:2009as}
  E.~J.~Copeland and M.~I.~Tsumagari,
  Phys.\ Rev.\ D {\bf 80} (2009) 025016
  [arXiv:0905.0125 [hep-th]].
\bibitem{magnus} W.Magnus, F. Oberhettinger and R. P. Soni, "Formulas and Theorems for the Special
Functions of Mathematical Physics", Springer-Verlag, Berlin, Heidelberg, New-York (1966).
\bibitem{colsys}
U. Ascher, J. Christiansen and R. D. Russell, Math. Comput. {\bf 33}
(1979), 659; ACM Trans. Math. Softw. {\bf 7} (1981), 209.
\bibitem{riedel}   B.~Hartmann, J.~Riedel and R.~Suciu, Work in preparation.
\end{thebibliography}
\end{document}